\documentclass[11pt,a4paper]{article}

\usepackage{mathptmx}       
\usepackage{helvet}         
\usepackage{courier}        
\usepackage{type1cm}        
\usepackage{makeidx}         
\usepackage{graphicx}        
\usepackage{multicol}        
\usepackage[bottom]{footmisc}

\usepackage{amsmath,amssymb,amsfonts}
\usepackage{color}

\def\note#1{\relax}
\definecolor{orange}{rgb}{1,0.5,0}

\newcommand{\hvii}{Henry VII}
\newcommand{\sping}{spin-glass}

\makeindex             

\begin{document}

\title{Community Detection in the Network of German Princes in 1225: a Case Study}
{\small
\author{S.R. Dahmen\footnote{Corresponding author. E-mail: silvio.dahmen@ufrgs.br}\\
Instituto de F\'{\i}sica da UFRGS,
91501--970 Porto Alegre Brazil\\
\\
A.L.C. Bazzan\\
Instituto de Inform\'atica da UFRGS,
91501--970 Porto Alegre Brazil\\
\\
R. Gramsch\\
Historisches Institut der
Universit\"at Jena,  F\"urstengraben 3,  07743 Jena Germany\\
}
}

\maketitle

\begin{abstract}
Many social networks exhibit some kind of community structure. In particular, in the context of historical research, clustering of different groups into warring or friendly factions
can lead to a better understanding of how conflicts arise, and if they could be avoided. 
In this work we study the crisis that started in 1225 when the Emperor of the Holy Roman Empire, Frederick II entered a conflict with his son Henry VII,
which almost led to a rupture and dissolution of the Empire. We use a spin-glass-based community detection algorithm to see how good this method is in detecting
this rift and compare the results with an analysis performed by one of the authors (Gramsch) using standard social balance theory applied to History.
\end{abstract}

\section{Introduction}
\label{sec:intro}




One of the main tasks in network theory is the detection of communities, that is nodes that cluster together according to some pre-defined criteria. The question whether or not a network can be partitioned in such a way
is not trivial and it is contingent on the question being asked. There are many criteria on how a community can be defined and detected (see \cite{Fortunato2010} for an extensive review on the subject). 
In the context of social networks in general and historical networks in particular, clustering can have far-reaching consequences, especially when cluster are involved in conflicts.  Under a sociological perspective, a natural way of grouping nodes is that of social balance theory, a model of human relationships that can be traced back to the works of F. Heider on cognitive dissonance theory \cite{Heider1946}. It is built upon the notion that, in a triad of nodes, if two of the nodes are positively related, their relation to the third party should match (in Sec.~\ref{sec:heider} we discuss this with more detail and generalizations thereof). The application of this idea to historical events follows naturally. But the main question is: is Heider's theory a
meaningful historical tool?
In order to answer this question from the perspective of Heider's  theory, one of the authors studied the conflict that arose between the years of 1225 and 1335 in what is roughly today's Germany~\cite{Gramsch2013}. 
This conflict pitted the Emperor Frederick II his heir, Henry VII,  over some disputes with the Pope. Based on Heider's theory, Gramsch showed that the dispute led to a rift among the prince-electors, some of which remained
behind the emperor while others backed his son. As this threatened the stability of the empire Frederick II had his son disavowed and imprisoned. 

One of the ultimate goals of historical research is to try to understand
how objectively events can be described, as we tend to analyze them with our preconceptions or worldviews. Medieval chroniclers were usually commandeered by this or that group of nobles, and were thus biased in their interpretation~\cite{Gramsch2013}. So, by trying to reduce the conflict to its bare essentials, one hopes to get rid of a subjective interpretation and pinpoint where intention ends and the power of structures, within which actors find themselves, begin.

The main goal of this paper is to use a clustering algorithm for this event and compare it to the results
found by Gramsch. Far from trying to rewrite history anew, since historical events
are extremely complex, spanning years and sometimes thousands of players, our goal is rather humble: to see if network analysis, particularly community  detection, may be used as a viable tool to help historian see patterns which otherwise could not be seen. 

This paper is organized as follows: we first give a brief overview of the event we are studying, the crisis of 1225 -- 1235 within the Holy Roman Empire caused by the Emperor Frederick II and his son and successor, Henry VII. 
In Section~\ref{sec:methods}, we present  materials and methods. 
We then discuss the results obtained by a traditional historical analysis and, after, we show how a \sping-based community  detection algorithm compares with that analysis.
\section{Background and Related Work}
\label{sec:bkg}

In the present work we deal with particular aspects of the coalition and conflicting forces that underlie the reign of \hvii\ in medieval Germany~\cite{Gramsch2013,Gramsch2014}.
It is common knowledge that in medieval times monarchic power was strongly restricted, and within the confines of the Holy Roman Empire particularly so. As a coalition of many sovereigns, a consensus among rulers was extremely important for a successful rule of the elected Emperor.
This became evident during the era of emperor Frederick II (1212--1250) and his son, king Henry VII (1220--1235). His  deposition by his father in 1235 was caused by the 
political incapacity of Henry, who gravely offended the princes and sacked them of their power.  Emperor Frederick had no choice but to disavow his son, lest he cause further damage 
to the authority of the Staufian dynasty. 


The conflict as a whole involved 68 sovereigns. The complexity of relationships between these is astounding and to reduce the dispute between Frederick
and Henry to a single question would be an oversimplification. However, as Gramsch convincingly demonstrates is his book, network analysis provides a new vista
on the overall structure of the network, which led to the deposition of Henry~\cite{Gramsch2013}.
%


\begin{figure}[h!]
\centering
\includegraphics[width=0.8\textwidth]{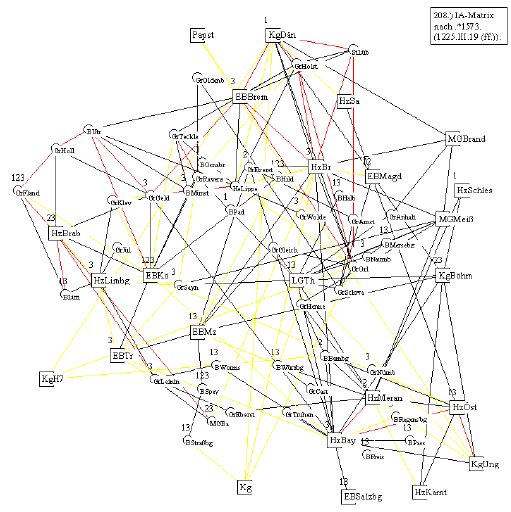}
\caption{The German political network from the socio-matrix of March 1225 (reproduction from Gramsch, 2013). For the meaning of some names, see below.}
\label{fig:g1225rough}
\end{figure}

In order to demonstrate the role of conflicts and coalitions that underlie the  historical event just described, Gramsch used clustering techniques~\cite{Gramsch2013,Gramsch2014}.
He depicted the political system of the medieval German empire as a network of princes, kings, counts, bishops and other sovereigns (who we shall generically call actors henceforth).
Based on Heider's structural balance theory \cite{Heider1946} (see Section~\ref{sec:heider}), he was able to characterize not only the existence of a relationship between actor (node) A and actor B (i.e., an edge in the network), but also that such relationships could be neutral, negative (hostile), or positive (friendly).
Positive relationships in this context can be kinship, political alliance, personal contact; whereas negative relationship are normally conflicts of various natures such as territorial or 
status competition, legal and military conflicts, etc. This analysis was carried out over a period of ten years of political relations and interactions among actors (from 1225 to 1235). 
These form the so-called socio-matrices, which can be identified with adjacency matrix, albeit with negative entries.

As mentioned previously, in his work Gramsch investigated a network composed of 68 actors, as well as the political relations and interactions between them over a period of ten years (from 1225 to 1235). These form the so-called socio-matrices, which can be identified with adjacency matrix, albeit with negative entries.
As in  \cite{Gramsch2013}, we use one socio-matrix for each year (unless  otherwise stated), i.e., the temporal dimension here is one year.
Such socio-matrices can be depicted as a network. For instance, Fig. \ref{fig:g1225rough} is a reproduction of Fig. 6 in  \cite{Gramsch2014}.
It shows friendly, hostile and neutral (ambivalent) relationships in black, red, and yellow respectively.

\begin{figure}[h]
\centering
\includegraphics[width=0.7\textwidth]{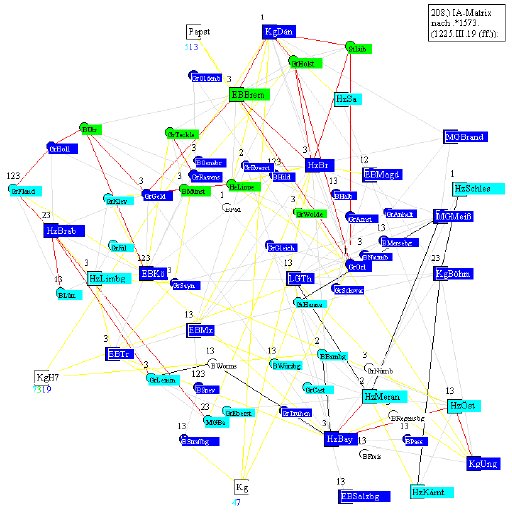}
\caption{The cluster structure of the German political network  of March 1225 (reproduction from Gramsch, 2013).}
\label{fig:g1225clusters}
\end{figure}

In order to cluster actors, Gramsch imposes that within a cluster there should be no negative connections (conflicts), without exception.
For the year 1125 this then translates into the clustering shown in Fig. \ref{fig:g1225clusters} (reproduction of Fig. 7 in \cite{Gramsch2014}).
He is also able to identify particular groups of actors such as the group of North Germany or the Lower Rhine.

His most important result is the 
detection of a dual structure in the network of princes in 1225 with the dominant blue cluster on the one hand and the turquoise and green clusters on the other hand. Both groups are separated by various conflicts.
We recall that,  previously, these conflicts were considered in isolation.
However,  \cite{Gramsch2014} showed that there were hidden relations between them.
For instance, in 1225, emperor Frederick II  predominantly collaborated with actors of the turquoise cluster while king Henry VII tended to form an alliance with blue actors. 
This then shows the origins of the  later conflict between the  father and the son.

Further, this analysis was able to show what happened between the years 1232 and 1235 (see figures in  \cite{Gramsch2014}), namely,
which actors stay together in one cluster, which ones have changed political coalitions, and how the front line of conflicts has changed geographically.
In short, one can observe that the political situation in 1232 was characterized by an antagonism of two  factions which are each composed of two clusters.
These two factions were, each, supported either by Frederick or Henry. Between 1232 and 1234, Frederick decided to depose his son in order to avoid further consequences
and recover the complete control over his empire.
These two antagonistic factions then start to decay in 1233 and disappear almost completely by 1235.

The investigation of network structures in history is not new. For example, in \cite{Padgett&Ansell1993}, Padgett and Ansell analyze the centralization
of political parties and elite networks that underlie the birth of the Renaissance state in Florence. In this study, some clustering techniques are also considered but the focus is on correlations of marriage, trade, partnership, bank, and real estate relations. Moreover, the techniques they used are not those employed in the present paper, so we concentrate on the work of Gramsch~\cite{Gramsch2013,Gramsch2014}

\section{Materials and Methods}
\label{sec:methods}

In this section we discuss the main methods used in our approach: Heider's structural balance theory and the Potts Model.
Following, we discuss their use for analyzing the network of 68 actors who take part in the historical event mentioned in Section~\ref{sec:bkg}.

\subsection{Heider's structural balance theory}
\label{sec:heider}

\begin{figure}
  \centering 
       \includegraphics[scale=0.4]{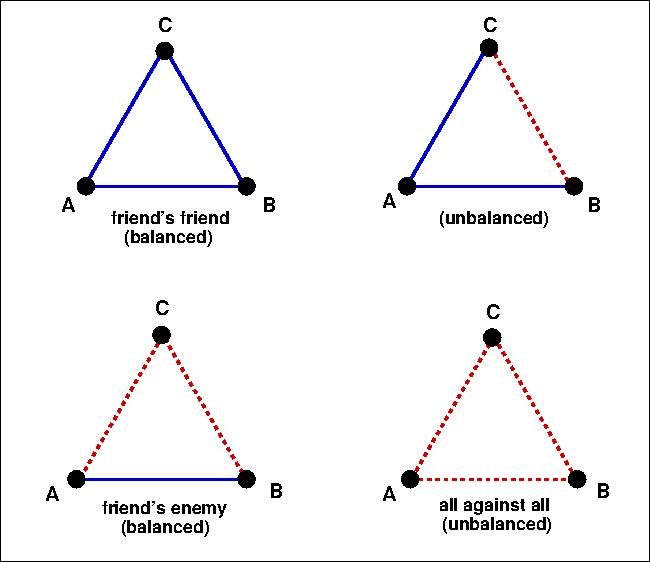}
        \caption{Cognitive balance according to F. Heider in the case of a triad of nodes and their mutual relations. A straight line depicts a positive relationship (friendship), while a broken line represents
a negative one (enmity). The column on the left depict the two possible balanced social relationships: my friend's friend/enemy is my friend/enemy. The column on the right depicts two unbalanced situations.}
        \label{fig:balance}
\end{figure}

In his seminal work of 1946 Heider asks the question of how an individual A's attitude towards B influence
the way a third individual C relates to B. This originated his structural balance theory, which basically states that a society is balanced when `a friend's friend (enemy) is also my friend (enemy)'.  If all triads of a network of relationships are balanced, the network is said to be balanced. There can be also situations where a positive relationship to an individual does not necessarily imply that his friend will also be positively related to that individual. Situations like this imply that the network is unbalanced. In Fig.~\ref{fig:balance} we depict the possible situations in a triad of relationships.
The question is whether a network of individuals with this kind of relationships can be grouped into separate communities.  Harary~\cite{Harary1953} showed that if a connected network is balanced, it can be split into two
opposing clusters.This was later generalized to cycles with more than $3$ individuals, to the idea of $k$-cycles~\cite{Davis1967,Cartwright&Harary1968}. 
A network is k-balanced if it can be divided into $k$ clusters where within each cluster there are only positive relationships.
In real life, however, not all clusters are balanced. Even if one is able to cluster nodes, there will always be within a cluster of positive relations some nodes with negative ones, as well as the opposite. The number of such misplaced  links is called 'frustration', a term  borrowed from the physics of spin systems: 
depending on the geometry and boundary conditions of the atomic lattice, it is impossible to have a pure antiferromagnetic state (of alternating spins). In network parlance, minimizing frustrations means the following: consider a network for which one can write down an adjacency matrix $A$, whose elements $A_{ij}$ are defined as follows:
\[
    A_{ij}= 
\begin{cases}
   1,& \text{if relation between node $i$ and node $j$ is positive } \\
   -1,              & \text{otherwise}
\end{cases}
\]
Frustration can be written as in Eq.~\ref{eq:frustration} below, where ${\sigma}$ represents a clustering and $\sigma_{i,j}$ means the cluster to which $i,j$ belong. $\delta$ is Kronecker's delta, with $\delta(\sigma_i,\sigma_j) = 1$ if $\sigma_i = \sigma_j$ and $0$ otherwise \cite{Traag&Bruggeman2009}.

\begin{equation}
 F({\sigma}) = - \sum_{ij} A_{ij}\delta(\sigma_i,\sigma_j)\; 
 \label{eq:frustration}
\end{equation}

The task is to find a configuration which minimizes this quantity. 
Frustration, in historical context,  means finding coalitions of actors where some may not be natural allies but nonetheless, based on some ulterior purpose, decide on joining forces. 
Due to this similarity between of Heider's ideas and methods in spin systems, it is natural to expect some methods from statistical mechanics of spin models will play a role in this scenario.

\subsection{Spin-Glass-Potts Model}


The resemblance between magnetic domains and clusters led Reichardt and Bornholdt to introduce a method of community detection based on a mapping between a graph and a q-state Potts Model~\cite{Reichardt+2006}. They considered only positive links between nodes, but
Traag and Bruggeman generalized it to account for the possibility of hostile links~\cite{Traag&Bruggeman2009}. We describe their method below.

The main idea is to reward positive links within a cluster and punish negative ones. At the same time
one wants to punish positive links between clusters and reward negative ones. One has a natural quality
function in the Hamiltonian of the system, {\it i.e.} its energy for a given configuration \{$\sigma$\} $=$ \{$\sigma_1, \sigma_2, \sigma_3,\cdots$\} of clusters $\sigma_1$, $\sigma_2$ etc. In order to write this Hamiltonian, one first breaks the adjacency matrix $A$ in two parts: 
\[
\begin{cases}
  A^{+}_{ij} = A_{ij}  & \text{if $A_{ij}>0$} \\
  A^{-}_{ij} = -A_{ij} & \text{if $A_{ij}<0$}
\end{cases}
\]
One rewards positive links between nodes $i$ and $j$ if they are in the same partition by a quantity $a_{ij}$ and penalize absent positive links when these nodes are in the same cluster by a quantity $b_{ij}$.
\begin{equation}
H^{+} ({\sigma } ) = \sum_{i,j}\left[ -a_{ij} A^{+}_{ij} +b_{ij}(1-A^{+}_{ij})\right]\delta (\sigma_i, \sigma_j ) 
\label{eq:h}
\end{equation}
In Eq.~\ref{eq:h} $\delta$ is Kronecker's delta function. By choosing $a_{ij} = 1- b_{ij}$ and $b_{ij} = \gamma^{+}p^{+}_{ij}$, where $p^{+}_{ij}$ is the probability that links $i$ and $j$ are positively connected, one regains the model of Reichardt and Bornholdt:
\begin{equation}
H^{+} ({\sigma } ) = - \sum_{i,j}( A^{+}_{ij} -\gamma^{+} p^{+}_{ij} )\delta (\sigma_i, \sigma_j ) 
\end{equation}
It is important to note that the choice of a probability distribution for $p_{ij}$ corresponds a the null model, relative to which clustering can be ascertained. $\gamma$ is a parameter used to
tune between the two competing terms. In an analogous way, one may define the negative part of the Hamiltonian,
which favors negative links between different clusters $\sigma_i$ and $\sigma_j$ and punishes positive ones:
\begin{equation}
H^{-} ({\sigma } ) = \sum_{i,j}( A^{-}_{ij} -\gamma^{-} p^{-}_{ij} )\delta (\sigma_i, \sigma_j )  
\end{equation}
where $\gamma^{-}$ and $P^{-}_{ij}$ have the analogous interpretation as in the positive case. By combining the two parts and remembering
that $A = A^{+} - A^{-}$, one can finally write
\begin{equation}
H ({\sigma } ) = -\sum_{i,j}\left[ A_{ij} -(\gamma^{+}p^{+}_{ij} - \gamma^{-}p^{-}_{ij} )\right]\delta (\sigma_i, \sigma_j ) 
\end{equation}
which is the quality function that has to be optimized. The choice of null model is important: the simplest one is to define the $p_{ij}$'s by
the proportion of positive links and negative links n the network relative to the total possible number of links, as in Eq.~\ref{eq:p}, where $m^{\pm}$ is the number of positive and negative links respectively.
\begin{equation}
p^{\pm}_{ij} = \frac{m^{\pm}}{n(n-1)}
\label{eq:p}
\end{equation}
 However in order to obtain a random model which maintain the degree
distribution of each vertex and thus come closer to the real network, one may define
\begin{equation}
p^{\pm}_{ij} = \frac{^{\pm}k_{i}\;^{\pm}k_{j}}{m^{\pm}}
\end{equation}
where $^{\pm}k_i$ represents the degree of positive (negative) links of a given node $i$. 
We refer the interested
reader to ~\cite{Traag&Bruggeman2009} where possible choices of parameters and their relation to Frustration and Modularity are explained. The optimal configuration can now be obtained by any heuristic method
from statistical mechanics. The routine we used is based on simulated annealing as explained in~\cite{Traag&Bruggeman2009}.

\subsection{Detecting Communities Using Spin-Glass}
In order to detect the community structure for the conflict between Frederick and his son, we used the igraph implementation of the \sping\ algorithm (Python variant). 
Each actor is represented by an abbreviated name, as indicated in Table~\ref{tab:actors}.
As in  \cite{Gramsch2013}, we use one socio-matrix (adjacency matrix) for each year (unless  otherwise stated), i.e., the temporal dimension here is one year.

We use a set of adjacency matrices (prepared by R. Gramsch, roughly one for each year), where $A_{ij}$ indicates whether or not there is a relationship between actors $i$ and $j$, and, if there is, whether it is neutral, friendly, or hostile.
We excluded, based on a suggestion of R. Gramsch,  all relationships  associated with Frederick II (Kg), Henry VII (KgH7) and the Pope (Papst), as these are the main actors of the conflict and served most of the time as liaisons between opposite groups. 
This way they biased the clustering and this hides important patterns.
Results reported in the next section, thus, do not include these three actors.
We remark that the same procedure was performed by Gramsch in his investigations; thus the results are comparable.

The \sping\ method needs
as input the number $n$ of communities. We chose $n=2$, to see whether the method would lead to a
partitioning of the network comparable to that found by Gramsch. If one gives a higher value of $n$, the method
will produce $n$ communities but normally for $n$ above a certain threshold (in some of our cases $5$ or above), the routine will give always at most 5 clusters, usually less.




\begin{table}
\centering
\caption{Some of the actors that appear in the depictions and how their names are shortened. The list
does not include all sovereigns.}
\begin{tabular}{ll|ll}
\hline 
Abbr. & Name & Abbr. & Name \\ \hline
Kg & Emperor Frederick II & KgH7 & King \hvii \\
HzBay & Duke of Bavaria &EBMz & Archbishop of Mainz \\
KgBoe & King of Bohemia & EBKoe & Archbishop of Cologne\\
BMuenst & Bishop of Muenster & KgUng & King of Hungary\\
Papst & Pope & GrHolst & Prince of Holstein\\
BUtr & Bishop of Utrecht & MGMeis & Margrave of Meissen\\
\hline
\end{tabular}
\label{tab:actors}
\end{table}

\section{Results}
\label{sec:results}

\begin{figure}
\centering
\includegraphics[width=0.8\textwidth]{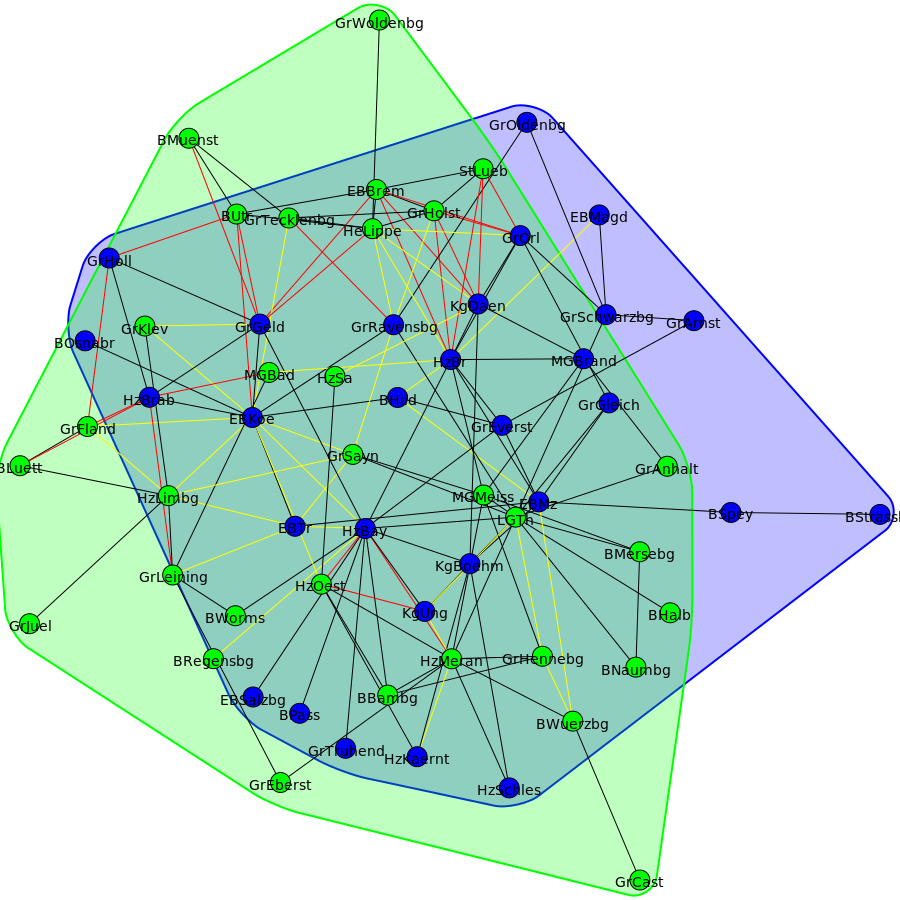}
\caption{The structure of the communities (clustering) -- year 1225}
\label{fig:1225nomain}
\end{figure}

\begin{figure}
\centering
\includegraphics[width=0.8\textwidth]{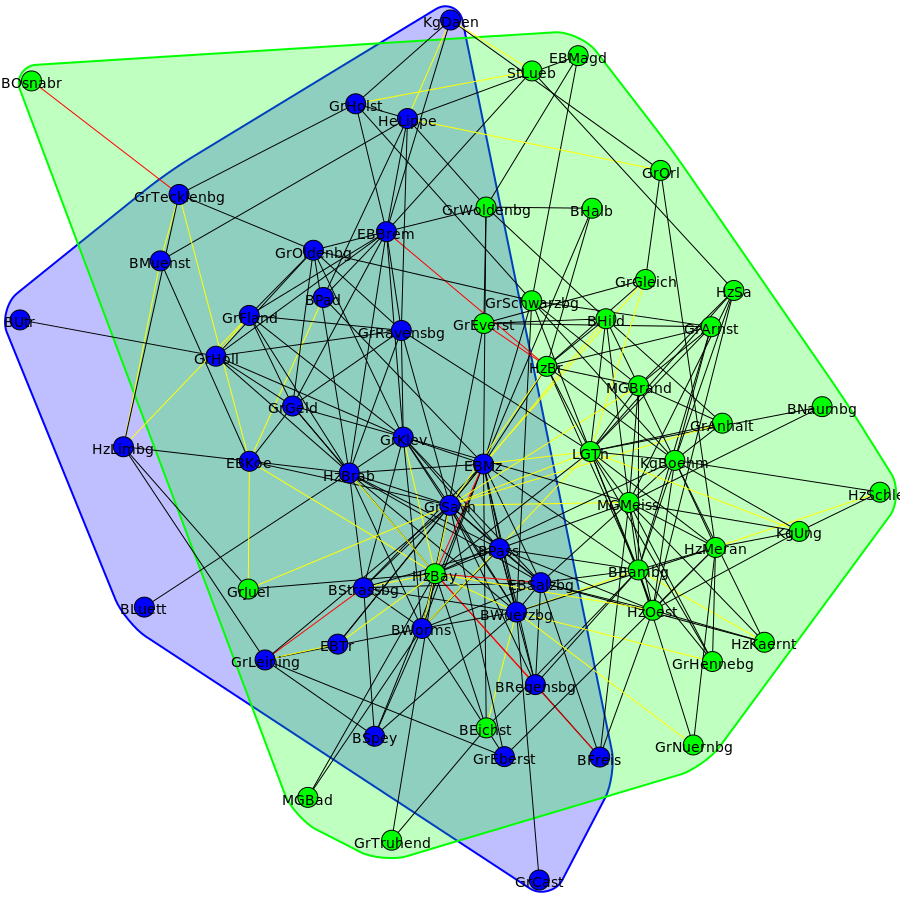}
\caption{The structure of the communities (clustering) -- year 1235}
\label{fig:1235nomain}
\end{figure}

We have run the \sping\ with, as mentioned, the number of spins set to 2, producing thus partitions that should separate the conflicting parties.
We did this for each year. Figures \ref{fig:1225nomain} and \ref{fig:1235nomain} show, for the sake of illustration, the clusterings for years 1225 and 1235 respectively\footnote{We remark that, obviously, this is the result of a single run, thus different runs can produce slightly different partitions.}.
Please notice the reduction of red edges (hostility) in the year 1235.

In order to compare the quality of the clustering produced originally by Gramsch in \cite{Gramsch2013} with those from the \sping\ method, we use the Rand index, which
is defined in the usual way as in Eq.~\ref{eq:ri}, where $a$ is the number of pairs of elements which are in the same set in both partitions $X$ and $Y$, and
$b$ is the number of pairs which are in different sets in partitions $X$ and continue to be so in $Y$. $n$ is the number of nodes. A Rand index of $1$ implies total agreement (sets are the same) while a $0$ implies total disagreement.

\begin{equation}
R=\frac{a+b}{\binom n2}
\label{eq:ri}
\end{equation}

Table~\ref{tab:ri} shows the Rand indexes when we do a comparison, year by year, with the original partitioning of Gramsch. 
We remark that, since the \sping\ method is not deterministic, 
we ran \sping\ community detection 30 times for each year.
Thus the table also shows the standard deviation associated with the mean value.

\begin{table}
\centering
\caption{Rand indexes (mean and standard deviation), by year}
\begin{tabular}{ccc|ccc}
\hline
Year & \multicolumn{2}{c|}{Rand index} & Year & \multicolumn{2}{c}{Rand index} \\
\cline{2-3}  \cline{5-6} 
& mean & st. dev. & & mean & std. dev. \\ \hline
1225 & 0.78 & 0.06 & 1226 & 0.8 & 0.06 \\
1227 & 0.66 & 0.15 & 1228 & 0.65 & 0.13\\
1229 & 0.73 & 0.05 & 1230 & 0.53 & 0.03\\
1231 & 0.84 & 0.9 & 1232 & 0.85 & 0.08 \\
1233 & 0.87 & 0.09 & 1234 & 0.78& 0.04\\
1235 & 0.87 & 0.04 & &\\
\hline
\end{tabular}
\label{tab:ri}
\end{table}

The values, as can be seen in the table, indicate a good agreement between the \sping method and Gramsch's original partitioning, based on Heider's structural balance. We would like to point out that, for the year
$1230$,
the agreement is comparatively low. This is due to the fact that in 1230 there occurred a temporary agreement between sovereigns. Quoting Gramsch (\cite{Gramsch2013}, p. 222): `During the first quarter of the  1230, when peace talks between the Emperor and the Pope began, the
sovereigns placed themselves in such a close [league] as it was never to be seen again: 58 joined
into one coalition.' So for this year there is only one cluster. Since the method requires a {\it
a priori} number of cluster to be created, which was set to $2$, the Rand index is smaller and is about
$0.5$, which corresponds to the probability of placing nodes with a 50-50 change on each cluster.

\section{Conclusion}
In this paper we applied a community detection algorithm to determine clusters of opposing sovereigns in conflict in medieval Germany, which took place between 1225 and 1235 and pitted the Emperor Frederick II against his son
Henry VII. We used a spin-glass-based algorithm to create clusters and to ascertain its feasibility as a tool in historical research, we compared the results with the partitioning previously done by one of the authors based on Heider's structural balance theory. For this we calculated the Rand index to compare partitions. Our results show good agreement with the historical method, from a minimum of 50\% in the worst case, as explained previously, to an agreement of 87\%. 

\section*{Acknowledgements}
Ana Bazzan is grateful to a CNPq grant. We thank Aline Weber for helping with the programming in Python.

\end{document}